\documentclass{optica-article}

\journal{opticajournal} % for journals or Optica Open

\articletype{Research Article}

\usepackage{lineno}
\usepackage{siunitx}

\begin{document}

\title{Circular photonic crystal grating design for charge-tunable quantum light sources in the telecom C-band}

\author{Chenxi Ma,\authormark{1} Jingzhong Yang,\authormark{1} Pengji Li,\authormark{1} Eddy P. Rugeramigabo,\authormark{1} Michael Zopf,\authormark{1,$\dag$} and Fei Ding\authormark{1,2,*}}

%\address{\authormark{1}Institut f{\"u}r Festk{\"o}rperphysik, Leibniz Universit{\"a}t Hannover, Appelstra{\ss}e 2, 30167 Hannover, Germany\\
\address{\authormark{1}Leibniz University Hannover, Institute of Solid State Physics, Appelstra{\ss}e 2, 30167 Hannover, Germany\\
%\authormark{2}Laboratorium f{\"u}r Nano- und Quantenengineering, Leibniz Universit{\"a}t Hannover, Schneiderberg 39, 30167 Hannover, Germany}
\authormark{2}Leibniz University Hannover, Laboratory of Nano and Quantum Engineering, Schneiderberg 39, 30167 Hannover, Germany}

\email{\authormark{$\dag$}michael.zopf@fkp.uni-hannover.de}
\email{\authormark{*}fei.ding@fkp.uni-hannover.de}
%% email address is required; see note below about the corresponding author designation

% use {asbstract*} to suppress the copyright line. Copyright information will be added in production

\begin{abstract*} 
Efficient generation of entangled photon pairs at telecom wavelengths is a key ingredient for long-range quantum networks. While embedding semiconductor quantum dots into hybrid circular Bragg gratings has proven effective, it conflicts with $p$-$i$-$n$ diode heterostructures which offer superior coherence. We propose and analyze hybrid circular photonic crystal gratings, incorporating air holes to facilitate charge carrier transport without compromising optical properties. Through numerical simulations, a broad cavity mode with a Purcell factor of 23 enhancing both exciton and biexciton transitions, and exceptional collection efficiency of \qty{92.4}{\percent} into an objective with numerical aperture of 0.7 are achieved. Furthermore, our design demonstrates direct coupling efficiency over \qty{90}{\percent} into a single-mode fiber over the entire telecom C-band. The hybrid circular photonic crystal grating thereby emerges as a promising solution for the efficient generation of highly coherent, polarization-entangled photon pairs.
\end{abstract*}

%%%%%%%%%%%%%%%%%%%%%%%%%%  body  %%%%%%%%%%%%%%%%%%%%%%%%%%
\section{Introduction}
Entangled photon pairs are essential building blocks for a quantum internet, enabling connections between distant nodes \cite{Kimble:2008} through entanglement-based quantum key distribution \cite{Yin:2020,Schimpf:2021} and quantum repeater schemes \cite{Simon:2007}. Although spontaneous parametric down-conversion (SPDC) sources have been the main workhorse for testing these scenarios, they face limitations in achieving high brightness without compromising the single-photon purity \cite{Senellart:2017}. In contrast, epitaxial semiconductor quantum dots (QDs) offer on-demand single-photon generation with low multi-photon probability and high brightness simultaneously. Notably, photons from QDs demonstrate high indistinguishability comparable with that generated from SPDC sources, particularly under resonance fluorescence (RF) and stimulated resonant two-photon excitation (TPE) schemes \cite{Somaschi:2016,Tomm:2021,Sbresny:2022}. Polarization-entangled photon pairs with fidelity up to 0.987 have been demonstrated via TPE of the biexciton state (XX) followed by a cascade decay via the exciton state (X) \cite{Schimpf:2021}. Recent progress has been achieved in improving the quality of QDs capable of emitting single photons or entangled photon pairs directly in the telecom C-band \cite{Zeuner:2021,Lettner:2021,Vajner:2023,Nawrath:2023,Joos:2023}. This advancement paves the way for utilizing optical fiber networks with minimal transmission loss and therefore increasing the achievable distances in quantum communication applications \cite{Yang:2023}. However, due to the high refractive index of the semiconductor host matrix, total internal reflection limits the fraction of photons extracted from QDs, necessitating their integration into photonic nanostructures such as weak-coupling optical microcavities \cite{Sapienza:2015,Somaschi:2016,Liu:2018:PhC} or photonic crystal waveguides \cite{Uppu:2020}. By harnessing the Purcell effect, the emitted photons can be efficiently funneled through the desired optical mode, resulting in high collection efficiency. In addition, the Purcell effect can enhance radiative decay rates of the excitonic transitions to improve the photon indistinguishability \cite{Purcell:1946,Varoutsis:2005}. Among various photonic nanostructures, hybrid circular Bragg gratings (CBGs) have demonstrated superior suitability for the generation of polarization-entangled photon pairs \cite{Wang:2019,Liu:2019,Rota:2022}. The broad cavity mode enhances the radiative decay rates of X and XX asymmetrically, thereby increasing their lifetime ratio and improving the photon indistinguishability \cite{Huber:2013,Scholl:2020}. Its broadband collection efficiency enhancement facilitates more relaxed requirements on the QD emission wavelength. Moreover, its planar structure allows for convenient integration onto micromachined piezoelectric actuators for the efficient transfer of anisotropic strain \cite{Trotta:2016,Chen:2016}. This configuration serves as a powerful tool for the simultaneous tuning of emission wavelength and maximization of the degree of entanglement by eliminating the exciton fine structure \cite{Hudson:2007,Rota:2022}.

Another challenge for QDs as quantum light sources, being embedded in a solid-state environment, is decoherence processes due to phonons, charge fluctuations, and spin fluctuations \cite{Lodahl:2015}. Highly coherent and blinking-free photon emission in the near-infrared range has been demonstrated by embedding QDs into a $p$-$i$-$n$ diode heterostructure to stabilize the charge environment via Coulomb blockade \cite{Zhai:2020,Zhai:2022}. However, applying this charge-tuning technique necessitates electrical contacts on both top and bottom doping layers for vertical biasing. Conventional CBGs featuring fully etched trenches are incompatible with this requirement. An apparent solution involves incorporating straight bridges for electrical connection \cite{Barbiero:2022,Singh:2022}. However, this introduces a waveguiding effect, reducing optical confinement and, consequently, the Purcell factor $F_\mathrm{P}$. Moreover, the discontinuous grating leads to a significant deviation from the desired Gaussian pattern in the far field, impairing prospective coupling efficiency into single-mode fibers (SMFs). These effects can be mitigated by narrowing the bridge such that the propagation mode is precluded. Nevertheless, this adjustment would undermine the effectiveness of charge control. Recently, a labyrinth geometry by shifting the position of the bridges between consecutive rings has been proposed to address these issues \cite{Buchinger:2023}. While this design partially restores the optical confinement and the quasi-Gaussian far-field pattern, it remains deficient in collecting photons. Therefore, a design that enables electrical connection while delivering comparable performance with conventional bridgeless CBGs is still missing.

In the first decade of this century, a structure called circular photonic crystal (CPC) was extensively studied, showcasing microcavities with isotropic photonic band gap and high quality factor $Q$ \cite{Horiuchi:2004,Scheuer:2004,Chang:2005}. Its continuous surface is conducive to electrical connections. Here, we revitalize the concept of CPC by combining its advantages with CBGs, replacing the trenches with air holes arranged in the sunflower geometry. We conduct numerical investigations of the InP-based hybrid circular photonic crystal grating (CPCG) structure while maintaining a broadband reflector comprising a SiO\textsubscript{2} layer and a gold mirror underneath. The azimuthal gaps between air holes serve as channels for charge carriers, rendering CPCG compatible with $p$-$i$-$n$ diode heterostructures. Design principles and optical properties are explored and assessed via finite-difference time-domain (FDTD) simulations. The optimized design yields a Purcell factor of 23 and collection efficiency of \qty{92.4}{\percent} into an objective with numerical aperture (NA) of 0.7. Additionally, we examine the impacts of QD displacement and orientation, demonstrating the robustness against such fabrication imperfections and the highly approximated circular symmetry of the CPCG. Finally, the maximum coupling efficiency of \qty{95}{\percent} is achieved for direct photon collection into a commercial SMF.

\section{Simulation method}
\begin{figure}[tb]
\centering
\includegraphics[width=4.556in]{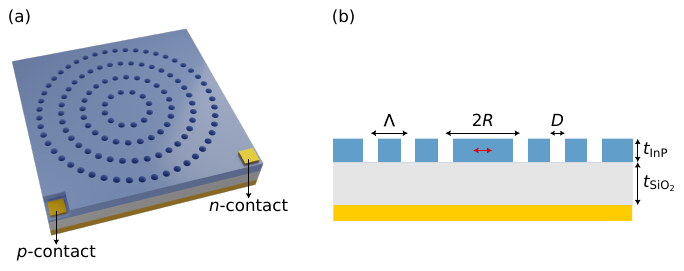}
\caption{Sketches of the hybrid CPCG device. (a) 3D illustration showing the InP layer with fully etched air holes in a sunflower-type geometry, on top of a SiO\textsubscript{2} layer and a gold mirror, including electrical contacts on both $p$- and $n$-layers. (b) Cross-sectional schematic with the relevant structure parameters, where $R$, $\Lambda$, and $D$ denote the central disk radius, the grating period, and the hole diameter, respectively.}
\label{fig:1}
\end{figure}

The optical properties of the CPCG are simulated using Ansys Lumerical FDTD 3D solutions. The thickness and the distance to the object of the perfectly matched layer boundaries are each at least half a wavelength in the corresponding medium to ensure sufficient absorption of incident light with minimal reflection. Because the optical properties are predominantly influenced by the CPCG region in the InP layer, its mesh size is set as \qty{20}{\nano\metre}, equivalent to over 22 grids per wavelength, to better mimic the circular shapes. For the remaining simulation region, we employ automatic non-uniform meshes to alleviate the demands on computational resources. To precisely simulate the device performance at cryogenic temperatures, which are crucial for minimizing the phonon thermal occupation and hence improving emission properties \cite{Reiter:2019}, refractive indices in the telecom C-band at \qty{4}{\K} ($n_{\mathrm{InP}}=3.135$, $n_\mathrm{SiO_2}=1.443$) are calculated or extracted from the literature, respectively \cite{Zielinska:2022,leviton:2006}. The emission from a QD is represented by an electric dipole oriented along the $x$ axis and positioned at the center of the InP layer, as pointed out by the red arrow in Fig. \ref{fig:1}(b). We define the extraction efficiency $\eta_{\mathrm{ext}}$ as the portion of the total dipole power entering the upper hemisphere, and the collection efficiency $\eta_{\mathrm{coll}}$ as the ratio of the integrated far-field power within the objective aperture angle to the total dipole power. The $\eta_{\mathrm{coll}}$ is calculated for NA = 0.7 throughout the paper unless specified otherwise.

\section{Results}

\subsection{Device design and parametric analysis}
The hybrid CPCG device illustrated in Fig. \ref{fig:1} incorporates an InP layer hosting InAs QDs at the center, situated atop a SiO\textsubscript{2} layer and a gold mirror. The InP layer thickness $t_\mathrm{InP}$ is set intentionally to \qty{300}{\nano\metre}, which is thicker than the critical thickness that only supports the fundamental transverse electric mode. This prevents degradation of QD emission properties caused by the $p$-type dopant diffusion \cite{Li:2018,Barbiero:2022}. The risk of coupling to the second-order mode is mitigated since the QD position corresponds to its node, rendering it unexcitable. The initial SiO\textsubscript{2} layer thickness $t_\mathrm{SiO_2}$ is chosen to be approximately half the wavelength in the medium, facilitating constructive interference between the upward emission and the reflected downward emission by the \qty{100}{\nano\metre}-thick gold mirror. Surrounding the dipole source, the sunflower-type CPCG consists of air holes etched through the InP layer, with their positions in the $xy$ plane given by \cite{Horiuchi:2004}
\begin{equation}
\begin{aligned}
x &= \left[R+\Lambda\left(N-1\right)\right] \cos \left(\frac{2m\pi}{nN}\right),\\
y &= \left[R+\Lambda\left(N-1\right)\right] \sin \left(\frac{2m\pi}{nN}\right),
\end{aligned}
\end{equation}
where $R$ denotes the radius of the central disk measured from the device center to the hole center in the first ring, $\Lambda$ represents the period of the concentric rings which satisfies the second-order Bragg condition, $N$ stands for the number of rings, while $n$ indicates $n$-fold rotational symmetry, and $m$ means the $m$th air hole in the $N$th ring. We set $n=12$ to maintain the degeneracy of two orthogonal fundamental cavity modes and to resemble a circular symmetry as closely as possible. As a consequence, the azimuthal distance between adjacent air holes is determined by the order of rotational symmetry and the ring radius, rather than being a controllable variable in the hole-CBG structure \cite{Jeon:2022}. Despite losing this degree of freedom, we prove in the following that the CPCG still yields good performance. Opting for a higher order rotational symmetry results in closer and smaller air holes, which is impractical as the gaps are too small to transport charge carriers, limited by the depletion width \cite{Berrier:2007}. Smaller holes also raise fabrication difficulties in the dry etching process for pattern transfer due to the increased aspect ratio.

\begin{figure}[tb]
\centering
\includegraphics[width=5.21in]{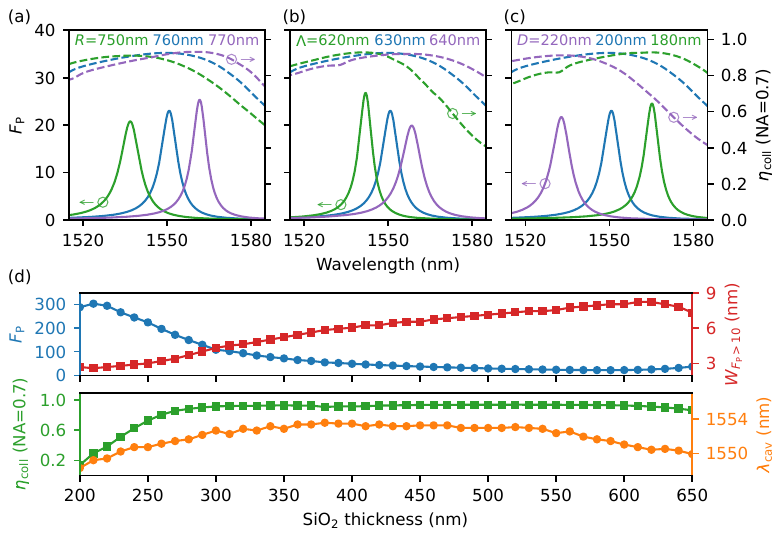}
\caption{Impacts of structure parameter variations on the optical properties of the CPCG. (a-c) Purcell factor $F_\mathrm{P}$ (solid) and collection efficiency $\eta_{\mathrm{coll}}$ (dashed) as a function of emission wavelength for different (a) $R$, (b) $\Lambda$, and (c) $D$. (d) Top panel: $F_\mathrm{P}$ and cavity spectral width $W_{F_\mathrm{P}>10}$ as a function of $t_\mathrm{SiO_2}$, showing that an increase in $F_\mathrm{P}$ is consistently accompanied with a reduction in $W_{F_\mathrm{P}>10}$, and vice versa. Bottom panel: $\eta_{\mathrm{coll}}$ and cavity resonance wavelength $\lambda_\mathrm{cav}$ as a function of $t_\mathrm{SiO_2}$, revealing their weak dependence on $t_\mathrm{SiO_2}$.}
\label{fig:2}
\end{figure}

To understand how structure parameters affect the device performance, we perform several groups of simulations where only one parameter is changed each time. In principle, the photonic crystal grating diffracts a portion of the emitted photons into a unidirectional vertical emission and reflects the remainder to establish the optical confinement. Therefore, the cavity mode characteristics are governed by the size of the central disk and the reflectivity of the grating. The radius of the disk $R$ is the most influential parameter affecting the cavity resonance wavelength $\lambda_\mathrm{cav}$, inducing a redshift when it increases, as shown in Fig. \ref{fig:2}(a). Concurrently, $F_\mathrm{P}$ and $\eta_{\mathrm{coll}}$ increase due to the proximity of the redshifted cavity mode to the upper band edge of the photonic crystal grating, where the reflectivity is slightly higher \cite{Liu:2019}. This enhances the optical confinement and the diffraction efficiency. Conversely, figure \ref{fig:2}(b) shows that an increase in the grating period $\Lambda$ weakens and broadens the cavity mode and flattens the $\eta_{\mathrm{coll}}$ curve. We attribute this to the reduced optical confinement for a fixed wavelength when the grating reflection spectrum redshifts. The hole diameter $D$ also plays an important role in controlling the device performance. When $D$ is larger, the effective refractive index of the disk is smaller, resulting in a blueshift of the cavity mode, as shown in Fig. \ref{fig:2}(c). Additionally, a larger $D$ intensifies the perturbation to the cavity mode, reducing the $Q$ factor by enhancing the radiation loss, which translates to favorable higher $\eta_{\mathrm{ext}}$ \cite{Akahane:2003}. However, $\eta_{\mathrm{coll}}$ is slightly reduced due to an enlarged emission half-angle, an effect that becomes more pronounced for smaller NA.

The SiO\textsubscript{2} layer thickness determines the optical path difference between the original upward emission and the reflected downward emission by the gold mirror. This parameter offers a convenient means to fine-tune the device performance with minimal impact on $\lambda_\mathrm{cav}$, as suggested by Fig. \ref{fig:2}(d). The maximum $F_\mathrm{P}$ of 303 is attained when $t_\mathrm{SiO_2} = \qty{210}{\nano\metre}$, albeit at the expense of greatly  suppressed $\eta_{\mathrm{coll}}$. One plausible explanation for this phenomenon is that, in this configuration, the optical path difference is approximately equal to $\lambda_\mathrm{cav}$. Accounting for the \ang{180} phase change upon reflection by the gold mirror, the reflected light destructively interferes with the original emission, impeding vertical emission from the QD. This can be equivalently regarded as an enhancement of the vertical contribution to the $Q$ factor, thereby leading to the observed maximum $F_\mathrm{P}$ \cite{Kim:2006}. This hypothesis is also supported by the fact that the cavity spectral width reaches its minimum at the same time. In order to achieve efficient photon collection, constructive interference needs to be satisfied. High $\eta_{\mathrm{coll}}$ exceeding \qty{90}{\percent} can be achieved over a broad range from \qtyrange{290}{630}{\nano\metre}, relaxing the constraints on $t_\mathrm{SiO_2}$. This flexibility also enables the replacement of the SiO\textsubscript{2} layer with a spin-coated indium tin oxide layer, which can serve as the bottom $p$-contact \cite{Sunde:2012}. Consequently, the need for precise control of the dry etching depth to expose the $p$-layer is alleviated, potentially enhancing the fabrication yield, especially for thin-film structures. A trade-off between $F_\mathrm{P}$ and $\eta_{\mathrm{coll}}$ can be tailored for various applications. For instance, $F_\mathrm{P}$ approaches 130 while $\eta_{\mathrm{coll}}$ remains above \qty{90}{\percent} when $t_\mathrm{SiO_2}=\qty{290}{\nano\metre}$, which is appealing for single-photon generation. As we focus on the generation of polarization-entangled photon pairs, a broad cavity mode is crucial for the simultaneous Purcell enhancement of both X and XX transitions and easier QD-cavity spectral coupling, improving optical properties and scalability. Therefore, we designate the optimal $t_\mathrm{SiO_2}$ as \qty{610}{\nano\metre}, achieving the broadest cavity mode without compromising $\eta_{\mathrm{coll}}$.

\subsection{Optimal device performance}
\begin{figure}[t]
\centering
\includegraphics[width=4.589in]{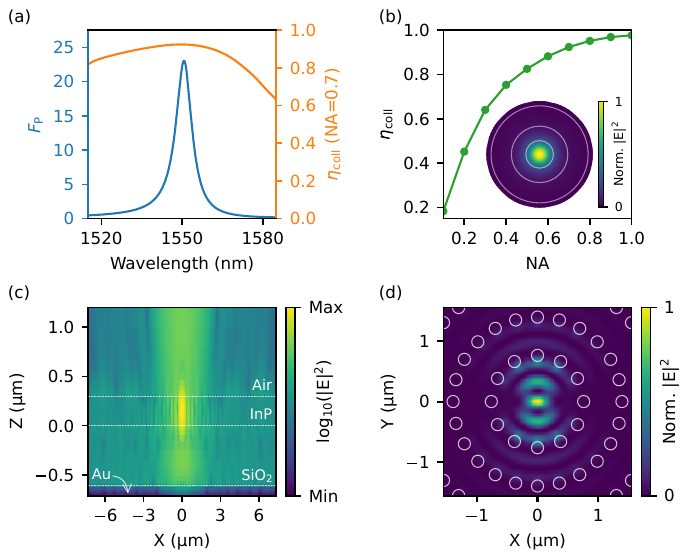}
\caption{Optical properties of the optimized CPCG structure. (a) $F_\mathrm{P}$ and $\eta_{\mathrm{coll}}$ as a function of emission wavelength. (b) NA-dependent $\eta_{\mathrm{coll}}$ at $\lambda_\mathrm{cav}$ showing efficient photon collection. The inset shows the highly converged far-field pattern, calculated by combining the far-field patterns of two orthogonal dipoles. The concentric circles and the boundary represent the acceptance angle for NA = 0.2, 0.4, 0.65, and 0.7, respectively. (c) Cross-sectional electric field intensity distribution in logarithmic scale showing the unidirectional emission. (d) In-plane intensity distribution of the cavity mode, with air holes represented by white circles.}
\label{fig:3}
\end{figure}

The preceding section highlights the ability to optimize the device performance by adjusting structure parameters. The final parameter set is chosen as $R = \qty{760}{\nano\metre}$, $\Lambda = \qty{630}{\nano\metre}$, $D = \qty{200}{\nano\metre}$, and $t_{\mathrm{SiO_2}} = \qty{610}{\nano\metre}$. Due to the smaller refractive index contrast in the photonic crystal grating, more periods ($N=10$) are necessary for sufficient optical confinement and efficient light extraction (see Fig. S1 of Supplement 1). This optimized device exhibits a broad cavity mode with a $F_\mathrm{P}$ of 23. For QDs emitting photons in the telecom C-band, the typical spectral separation between X and XX is around \qty{6}{\nano\metre} \cite{Zeuner:2021,Lettner:2021,Vajner:2023}. In our simulation, when XX is in resonance with the cavity, X experiences a $F_\mathrm{P}$ of roughly 6.8. Applying such asymmetric Purcell enhancement to the state-of-the-art lifetimes measured under TPE predicts a theoretical photon indistinguishability of \qty{92.9}{\percent} \cite{Vajner:2023,Scholl:2020}. In addition, considering the measured statistics of lifetimes and coherence times of X and XX of InAs/InP QDs grown in the droplet epitaxy mode, $F_\mathrm{P}>10$ is vital to realize transform-limited emission from both transitions \cite{Anderson:2021}. The corresponding cavity spectral width $W_{F_\mathrm{P}>10}$ is \qty{8.22}{\nano\metre}, which is broad enough to enhance the X and XX transitions simultaneously. Thanks to the broadband nature of the grating structure, maximum $\eta_{\mathrm{coll}}$ of \qty{92.4}{\percent} is achieved while exceeding \qty{80}{\percent} over a spectral range of \qty{60}{\nano\metre} from \qtyrange{1513}{1573}{\nano\metre}, as shown in Fig. \ref{fig:3}(a). This experimentally advantageous feature ensures efficient photon collection even when QDs are not in resonance with the cavity.

It is noteworthy that the effective refractive index of the air-hole ring gradually decreases, from the innermost to the outermost ring. This characteristic suppresses higher-order diffraction, contributing to a highly converged unidirectional far-field pattern, evident in the inset of Fig. \ref{fig:3}(b) and the cross-sectional electric field intensity distribution in Fig. \ref{fig:3}(c). As a result, high $\eta_{\mathrm{coll}}$ persists for objectives with small NA, achieving \qty{75.4}{\percent} for NA = 0.4, as plotted in Fig. \ref{fig:3}(b). This feature makes CPCGs attractive for employing cost-effective objectives or direct fiber coupling. The in-plane intensity distribution of the cavity mode, illustrated in Fig. \ref{fig:3}(d), reveals the interaction between the QD and the CPCG. The residual field intensity upon encountering the first ring clarifies the strong correlation between $\lambda_\mathrm{cav}$ and $R$. As discussed earlier, substituting the complete trench with air holes reduces the interaction area, resulting in a gentler change of the electric field at the cavity edges and, consequently, a slightly higher $Q$ factor in relative to CBGs \cite{Akahane:2003}. Superior results may be achieved using advanced optimization algorithms \cite{Bremer:2022,Rickert:2023}, including the incorporation of additional structure parameters such as chirping the grating period or apodizing the hole diameter, which remain interesting for future studies.

\subsection{Robustness against dipole displacement and finite rotational symmetry}
QDs are randomly distributed across the wafer due to the inherent randomness of the growth mechanism, posing a challenge to their integration into photonic nanostructures, as the performance relies heavily on the spatial coupling between the QD and the cavity. While fabricating arrays of multiple devices on high-density QD samples is an approach, it may compromise the single-photon purity due to nonresonant feeding if more than one QD is present \cite{Laucht:2010}. Therefore, deterministic fabrication of photonic nanostructures on low-density QD samples emerges as the most promising solution. Optical positioning techniques, especially the wide-field photoluminescence imaging for semiconductor QDs in the near-infrared region, have enabled deterministic device fabrication with positioning accuracy mainly limited by the electron beam lithography (EBL) alignment uncertainty \cite{Liu:2021}. However, the accuracy diminishes in the telecom range due to the dimer emission of telecom QDs and the inefficiency of InGaAs image sensors used in telecom cameras. To our best knowledge, the cutting-edge deterministic device fabrication in the telecom C-band achieves median overall cavity placement accuracy of approximately \qty{130}{\nano\metre} \cite{Holewa:2023}. Thus, it is crucial to investigate how the optical properties of CPCGs change when the QD is displaced from the cavity center with a similar level of mismatch.

To quantify the degradation of device performance caused by QD displacement, we move the dipole along $x$ and $y$ axes in steps of \qty{50}{\nano\metre}. When the dipole is moved along the $x$ axis, aligned with its orientation, the most obvious characteristics are the preserved maximum $\eta_{\mathrm{coll}}$ and the retained broad bandwidth. At the largest displacement of \qty{150}{\nano\metre}, $F_\mathrm{P}$ decreases to 11, corresponding to \qty{50}{\percent} of its original value. On the other hand, the cavity mode diminishes rapidly when the dipole is displaced along the $y$ axis. The reduction in $\eta_{\mathrm{coll}}$ is only \qty{5}{\percent}, but there is significant degradation in the bandwidth. These findings suggest that when a QD is not at the center of the CPCG, its horizontal and vertical linearly polarized dipoles undergo different levels of Purcell enhancement. The difference in radiative decay rates unbalances the branching ratio of the XX-X cascade, leading to a weakened entanglement \cite{Larqué:2009}. Nevertheless, the polarization selectivity induced by the QD displacement is advantageous for generating single-photon streams with a high degree of linear polarization, such as the previous demonstration of elliptical microcavities under RF that surpasses the typical \qty{50}{\percent} brightness limit imposed by the cross-polarization filtering scheme\cite{Wang:2019:Elliptical}.

\begin{figure}[t]
\centering
\includegraphics[width=5.208in]{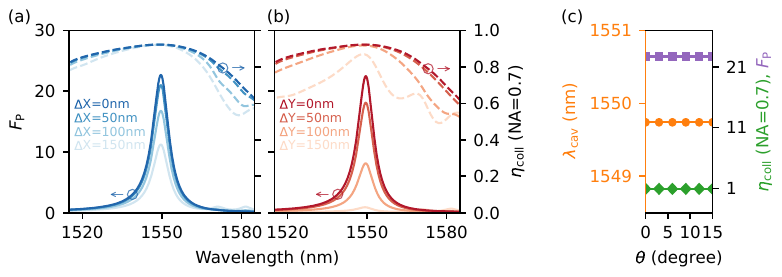}
\caption{Impacts of dipole displacement and orientation on the optical properties of the CPCG. (a, b) $F_\mathrm{P}$ and $\eta_{\mathrm{coll}}$ as a function of emission wavelength for different dipole displacement along (a) the $x$ axis and (b) the $y$ axis, showing stronger dependence along the $y$ axis for an $x$-oriented dipole. (c) $\lambda_\mathrm{cav}$, and corresponding $F_\mathrm{P}$ and $\eta_{\mathrm{coll}}$ as a function of dipole orientation $\theta$, demonstrating the circular symmetry of the CPCG.}
\label{fig:4}
\end{figure}

It is well known that the CPC possesses an isotropic photonic band gap \cite{Horiuchi:2004}. However, this feature does not guarantee that arbitrarily oriented dipoles will excite identical cavity modes. In order to assess the degree of circular symmetry of the CPCG, the dipole orientation $\theta$ is rotated from \ang{0} to \ang{15} with increments of \ang{3}, covering half the angle interval between adjacent symmetry axes. Interestingly, three key figures of merit, namely $\lambda_\mathrm{cav}$, $F_\mathrm{P}$, and $\eta_{\mathrm{coll}}$, exhibit no correlation with $\theta$. These results indicate that CPCGs with 12-fold rotational symmetry can be considered to possess circular symmetry, eliminating the last potential obstacle for it to replace CBGs.

\subsection{Direct coupling into single-mode fibers} \label{Direct coupling into single-mode fibers}

\begin{figure}[t]
\centering
\includegraphics[width=3.904in]{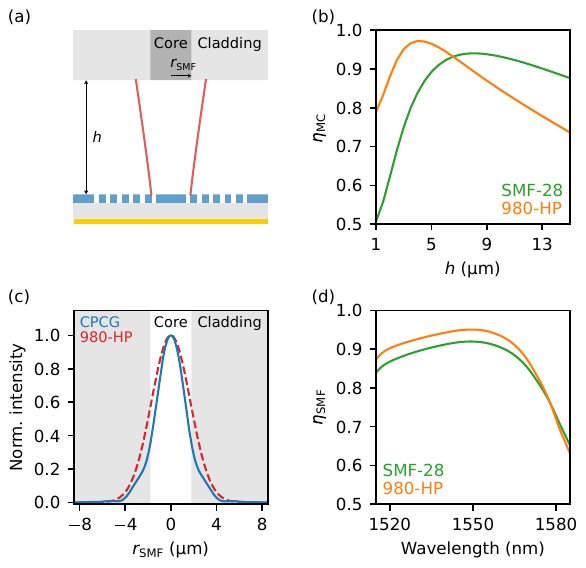}
\caption{Direct coupling of the emission from CPCGs into SMFs. (a) Cross-sectional schematic illustrating the direct fiber coupling configuration, with the MFD of the far-field pattern denoted by the red line. (b) $\eta_{\mathrm{MC}}$ of the optimized CPCG at $\lambda_\mathrm{cav}$ when coupling into SMF-28 and 980-HP fibers as a function of $h$. (c) Intensity profile lineshapes showing good agreement between the far-field pattern of the optimized CPCG and the Gaussian function of the propagation mode of a 980-HP fiber on the fiber facet at $h=\qty{4}{\micro\metre}$. (d) $\eta_{\mathrm{SMF}}$ when coupling into SMF-28 and 980-HP fibers at their respective optimal $h$ as a function of emission wavelength.}
\label{fig:5}
\end{figure}

For real-world applications, portable integrated quantum light sources are highly desirable. Direct coupling into optical fibers is particularly attractive as it eliminates the need for bulky optics, thereby allowing for significant miniaturization of the footprint. Moreover, while an objective is limited to collecting signals from a finite field of view, direct fiber coupling enables scaling up to access multiple emitters on the same chip. Two commercial SMFs commonly deployed for telecom applications are considered, namely SMF-28 and 980-HP with mode field diameters (MFDs) of \qty{10.4}{\micro\metre} and \qty{6.8}{\micro\metre}, respectively. The MFD represents the diameter where the mode intensity decreases to $1/e^2$ of its peak value. The normalized fundamental fiber mode amplitude $E_{\mathrm{SMF}}$ can be approximated by a Gaussian function \cite{Neumann:2013}:
\begin{equation}
%E_{\mathrm{SMF}} \approx \exp \left[-\frac{4\left(x^2+y^2\right)}{\left(\mathrm{MFD}\right)^2}\right].
E_{\mathrm{SMF}} \approx \exp \left(-\frac{r_{\mathrm{SMF}}^2}{w^2}\right),
\end{equation}
where $r_{\mathrm{SMF}}=\sqrt{x^2+y^2}$ is the radial distance from the beam center, and the beam radius is denoted by $w=\mathrm{MFD}/2$. We calculate the far-field amplitude of circularly polarized photons emitted from the XX-X cascade by combining the far-field patterns of two orthogonal dipoles, denoted by $E_{\mathrm{far}}$, to mimic the actual photon emission. To evaluate the compatibility of the CPCG with direct fiber coupling applications, we calculate the mode coupling efficiency $\eta_{\mathrm{MC}}$ between the far-field pattern of a CPCG and the fundamental mode of an SMF using the overlap integral method, as described by \cite{Neumann:2013}
\begin{equation}
\eta_{\mathrm{MC}}=\frac{\left|\int E_{\mathrm{far}} E_{\mathrm{SMF}}^* {\mathrm{d}}A\right|^2}{\int \left|E_{\mathrm{far}}\right|^2 {\mathrm{d}}A \int \left|E_{\mathrm{SMF}}\right|^2 {\mathrm{d}}A},
\end{equation}
where $E_{\mathrm{SMF}}^*$ is the conjugate of the Gaussian function approximating the fundamental mode amplitude on the fiber facet, and the integration area $A$ is a square with side lengths of 10 standard deviations of $E_{\mathrm{SMF}}$ to ensure maximal inclusion of the electric field. As the QD emission diverges while propagating, we project the far-field amplitude onto planes at different heights $h$, which represents the distance from the top surface of the InP layer to the fiber facet, as illustrated in Fig. \ref{fig:5}(a). Subsequently, we calculate the corresponding $\eta_{\mathrm{MC}}$. The lineshape comparison in Fig. \ref{fig:5}(c) demonstrates that the far-field pattern of the CPCG matches well with the fundamental mode of the 980-HP fiber, yielding maximum $\eta_{\mathrm{MC}}$ of \qty{97.2}{\percent} at $h=\qty{4}{\micro\metre}$, as shown in Fig. \ref{fig:5}(b). Given the larger core diameter and thus a larger MFD of the SMF-28 fiber, a slightly larger distance is required, achieving $\eta_{\mathrm{MC}}$ of \qty{94.1}{\percent} at $h=\qty{8}{\micro\metre}$. This high $\eta_{\mathrm{MC}}$ means that it is unnecessary to employ ultra-high NA fibers, as splicing them into SMF-28 fibers introduces additional loss and technical complexity. However, it should be noted that $\eta_{\mathrm{MC}}$ is sensitive to $h$, decreasing rapidly when the fiber drifts away from the optimal location. Other misalignments such as lateral offset or fiber tilt also induce degradation, and their impacts have been discussed in \cite{Schwab:2022}. These issues can be resolved by using adhesives to establish stable integration between the photonic chip and the fiber, in which case the structure parameters need to be adjusted to restore the desired optical properties due to the change in the refractive index contrast. In order to study the wavelength-dependent performance, we calculate the overall SMF coupling efficiency $\eta_{\mathrm{SMF}}$, defined as the product of $\eta_{\mathrm{ext}}$ and $\eta_{\mathrm{MC}}$, quantifying the ratio of the power entering and propagating in the fiber to the total dipole power. It can be observed that for both fibers at their respective optimal $h$, the broadband characteristic is preserved, as shown in Fig. \ref{fig:5}(d). For direct coupling into a 980-HP fiber, $\eta_{\mathrm{SMF}}$ reaches \qty{95}{\percent} and remains above \qty{91.1}{\percent} over the entire telecom C-band, surpassing the collection efficiency into an objective with NA = 0.7. Coupling into an SMF-28 fiber yields a maximum of \qty{92}{\percent} and a bandwidth of \qty{29}{\nano\metre} for $\eta_{\mathrm{SMF}}>\qty{90}{\percent}$. These results underscore the practical utility of the CPCG for direct fiber coupling applications.

\section{Conclusion}
In summary, we performed FDTD simulations to numerically study a CPCG design tailored for charge-tunable semiconductor QD-based single-photon or entangled photon-pair sources. By replacing the fully etched trenches in the CBG with air holes, the CPCG allows for electrical access to QDs without deteriorating the device performance. It outperforms alternative proposals such as the labyrinth geometry by achieving collection efficiency of \qty{92.4}{\percent} into an objective with NA = 0.7 and a more Gaussian-like far-field pattern. The optimized cavity mode, featuring a Purcell factor of 23, is sufficiently broad to offer improved photon indistinguishability via the asymmetric Purcell enhancement of exciton and biexciton transitions. We conclude that the performance of the CPCG remains unaffected by the in-plane QD orientation, and that they are robust against lateral QD displacement in single-photon generation scenarios. Furthermore, direct fiber coupling can be achieved with near-unity efficiency. The hybrid design allows for integration onto piezoelectric platforms, enabling dynamic strain-tuning of the emitter-cavity system for wavelength tuning and the elimination of exciton fine structure. The proposed design therefore enables a combination of charge-tuning and strain-tuning toward the development of application-ready quantum light sources emitting bright and indistinguishable photons with high entanglement fidelity. This capability is instrumental for global quantum communication through existing optical fiber networks.

\begin{backmatter}
\bmsection{Funding}
The authors gratefully acknowledge the German Federal Ministry of Education and Research (BMBF) within the projects QR.X (16KISQ015), SemIQON (13N16291), SQuaD (16KISQ117), and QVLS-iLabs: Dip-QT (03ZU1209DD), the European Research Council (QD-NOMS - No. GA715770, MiNet – No. GA101043851), MWK Niedersachsen (QuanTec - 76251-1009/2021), and the Deutsche Forschungsgemeinschaft (DFG, German Research Foundation) within the project InterSync (GZ: INST 187/880-1 AOBJ: 683478), and under Germany’s Excellence Strategy (EXC-2123) Quantum Frontiers (390837967). P.L. acknowledges the China Scholarship Council (CSC201807040076).

\bmsection{Acknowledgments}
The authors thank Leonardo Midolo and Frederik Benthin for fruitful discussions. C.M. is grateful to Peter Lodahl and the Rosenfeld Foundation for supporting the research visit at the Niels Bohr Institute, University of Copenhagen.
%The publication of this article was funded by the Open Access Fund of the Leibniz Universit{\"a}t Hannover.

\bmsection{Author contributions}
C.M. conducted the numerical simulations. J.Y. assisted in the simulations and the fiber mode coupling efficiency calculations. P.L. helped with the refinement of the working principle description of the CPCG. E.P.R. and M.Z. supervised the simulations. C.M. performed the data analysis and interpretation, and wrote the manuscript with input from all co-authors. F.D. supervised the project.

\bmsection{Disclosures}
The authors declare no conflicts of interest.

\bmsection{Data availability} Data underlying the results presented in this paper are not publicly available at this time but may be obtained from the authors upon reasonable request.

\bmsection{Supplemental document}
See Supplement 1 for supporting content. 

\end{backmatter}

%%%%%%%%%%%%%%%%%%%%%%% References %%%%%%%%%%%%%%%%%%%%%%%%%
%%%%%%%%%% If using BibTeX:
\bibliography{ref_main}

\end{document}

\maketitle

\section{Benchmark of the CPCG performance}
Table \ref{tab1:comparison} summarizes the optical properties of state-of-the-art circular Bragg grating (CBG) and CPCG designs that are compatible with electrical access. Our CPCG design exhibits the best performance in terms of overall collection efficiency $\eta_{\mathrm{coll}}$ and direct single-mode fiber coupling efficiency $\eta_{\mathrm{SMF}}$, which result from a more Gaussian-like far-field pattern.

\begin{table}[h]
\centering
\caption{\bf Comparison of CPCG performance with state-of-the-art electrically-tunable CBGs}
\begin{threeparttable}
\begin{tabulary}{\textwidth}{CCCCCCCCC}
\hline
Ref. & Device & Material & Wavelength (nm) & $F_\mathrm{P}$ & NA\tnote{$a$} & $\eta_{\mathrm{coll}}$ & $\eta_{\mathrm{SMF}}$ \\
\hline
\cite{Blokhin:2021} & CBG half–micropillar & GaAs & 1300 & 5 & 0.7 & \qty{75}{\percent} & \qty{22}{\percent}\tnote{$b$} \\
\cite{Schall:2021} & CBG on back-side DBR & GaAs & 925 & - & 0.8 & \qty{24.4}{\percent} & - \\
\cite{Barbiero:2022} & Hybrid CBG 4-WG & InP & 1550 & 20 & 0.65 & \qty{70}{\percent} & - \\
\cite{Buchinger:2023} & Hybrid CBG 4-LW & GaAs & 910 & 18 & 0.64 & \qty{59}{\percent} & $\sim$\qty{85}{\percent}\tnote{$c$} \\
\cite{Buchinger:2023} & Hybrid CBG 4-SW & GaAs & 910 & 17 & 0.64 & \qty{59}{\percent} & $\sim$\qty{82}{\percent}\tnote{$c$} \\
\cite{Shih:2023} & Hybrid CBG 4-LW & GaAs & 1550 & 29 & 0.8 & \qty{88}{\percent} & - \\
This work & Hybrid CPCG & InP & 1550 & 22 & 0.7 & \qty{92.4}{\percent} & \qty{95}{\percent}\tnote{$b$} \\
\hline
\end{tabulary}
\begin{tablenotes}
    \item[$a$] Numerical aperture
    \item[$b$] Coupling into 980-HP fibers
    \item[$c$] Overlap with fitted circular 2D Gaussian profiles, not exactly $\eta_{\mathrm{SMF}}$
\end{tablenotes}
\end{threeparttable}
\label{tab1:comparison}
\end{table}

\section{Impacts of the number of rings on CPCG performance}
\begin{figure}[ht]
\centering
\includegraphics[width=3.554in]{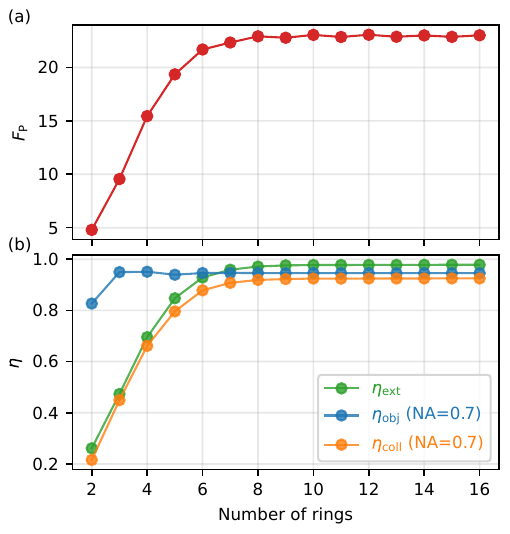}
\caption{Impacts of the number of rings on the optical properties of the CPCG. (a) Purcell factor $F_\mathrm{P}$ as a function of the number of rings. (b) Extraction efficiency $\eta_{\mathrm{ext}}$, and objective collection efficiency $\eta_{\mathrm{obj}}$ and $\eta_{\mathrm{coll}}$ for NA = 0.7 as a function of the number of rings.}
\label{fig:s1}
\end{figure}

Figure \ref{fig:s1} shows that an increase in the number of rings yields higher Purcell factor $F_\mathrm{P}$ and $\eta_{\mathrm{coll}}$. To facilitate a better interpretation of the results, a new figure of merit, namely the objective collection efficiency $\eta_{\mathrm{obj}}$, is introduced. This parameter is defined as the ratio of the integrated far-field power within the objective aperture angle to the dipole power that enters the upper hemisphere. In other words, $\eta_{\mathrm{coll}}$ is the product of extraction efficiency $\eta_{\mathrm{ext}}$ and $\eta_{\mathrm{obj}}$. It can be observed in Fig. \ref{fig:s1}(b) that the $\eta_{\mathrm{obj}}$ saturates rapidly above 3 rings, indicating that the far-field pattern has already converged. The primary limiting factor for $\eta_{\mathrm{coll}}$ is the low $\eta_{\mathrm{ext}}$, which can be attributed to stronger in-plane loss due to the smaller refractive index contrast in the photonic crystal grating, suggesting a slightly larger in-plane mode volume compared with CBGs \cite{Jeon:2022}. Hence, more rings are necessary for efficient photon collection. Both $F_\mathrm{P}$ and $\eta_{\mathrm{coll}}$ almost saturate at 8 rings, and reach complete saturation above 10 rings.

\begin{figure}[!h]
\centering
\includegraphics[width=3.188in]{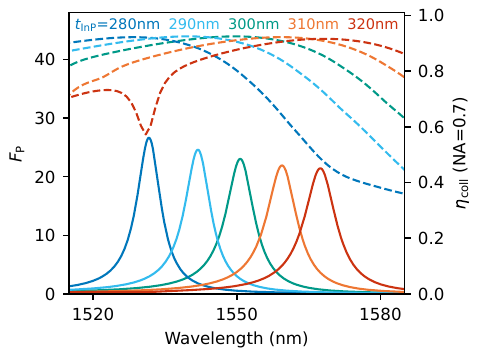}
\caption{Impacts of InP layer thickness $t_{\mathrm{InP}}$ on the optical properties of the CPCG. $F_\mathrm{P}$ (solid) and $\eta_{\mathrm{coll}}$ (dashed) as a function of emission wavelength for different $t_{\mathrm{InP}}$.}
\label{fig:s2}
\end{figure}

\section{Impacts of InP layer thickness on CPCG performance}
Figure \ref{fig:s2} demonstrates that with each \qty{10}{\nano\metre} reduction in the InP layer thickness $t_{\mathrm{InP}}$, the cavity resonance wavelength blueshifts by approximately \qty{10}{\nano\metre} as well. Simultaneously, there is an increase in $F_\mathrm{P}$ and a decrease in the full width at half maximum, indicating a higher quality factor $Q$. The $\eta_{\mathrm{coll}}$ remains unaffected by $t_{\mathrm{InP}}$. These results suggest that the thinnest possible InP layer should be employed, provided that the QD optical properties are not compromised.

\section{Different CPCG configuration for single-photon generation}
\begin{figure}[!h]
\centering
\includegraphics[width=4.58in]{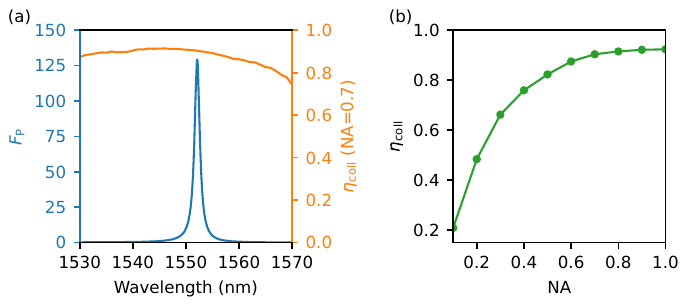}
\caption{Impacts of SiO\textsubscript{2} layer thickness $t_\mathrm{SiO_2}$ on the optical properties of the CPCG. (a) $F_\mathrm{P}$ and $\eta_{\mathrm{coll}}$ as a function of emission wavelength. (b) NA-dependent $\eta_{\mathrm{coll}}$ at $\lambda_\mathrm{cav}$ showing efficient photon collection.}
\label{fig:s3}
\end{figure}

As discussed in the main text, a trade-off between $F_\mathrm{P}$ and $\eta_{\mathrm{coll}}$ can be tailored for various applications by changing the SiO\textsubscript{2} layer thickness $t_\mathrm{SiO_2}$. At $t_\mathrm{SiO_2}=\qty{290}{\nano\metre}$, $F_\mathrm{P}$ reaches 129 while ensuring a $\eta_{\mathrm{coll}}$ above \qty{90}{\percent}, as shown in Fig. \ref{fig:s3}(a). The high $F_\mathrm{P}$ can greatly improve the optical properties of the single-photon emission. Figure \ref{fig:s3}(b) shows that the $\eta_{\mathrm{coll}}$ at different NA are slightly lower than the CPCG with a \qty{610}{\nano\metre} SiO\textsubscript{2} layer due to the incomplete constructive interference.

% Bibliography
\bibliography{ref_supplementary}